\documentclass[useAMS,usenatbib]{mnras}

\usepackage{url,times,hyperref,graphicx,amsmath,amsfonts,amssymb,aas_macros,color,epsfig,epstopdf,lipsum}
\usepackage{dblfloatfix} 

\newcommand{\hMpc}{{\ifmmode{h^{-1}{\rm Mpc}}\else{$h^{-1}$Mpc}\fi}}
\newcommand{\hkpc}{{\ifmmode{h^{-1}{\rm kpc}}\else{$h^{-1}$kpc}\fi}}
\newcommand{\hMsun}{{\ifmmode{h^{-1}{\rm {M_{\odot}}}}\else{$h^{-1}{\rm{M_{\odot}}}$}\fi}}
\newcommand{\ltsima}{$\; \buildrel < \over \sim \;$}
\newcommand{\gtsima}{$\; \buildrel > \over \sim \;$}
\newcommand{\lsim}{\lower.5ex\hbox{\ltsima}}
\newcommand{\gsim}{\lower.5ex\hbox{\gtsima}}

\def\lesssim{\mathrel{\hbox{\rlap{\hbox{\lower4pt\hbox{$\sim$}}}\hbox{$<$}}}}
\def\gtrsim{\mathrel{\hbox{\rlap{\hbox{\lower4pt\hbox{$\sim$}}}\hbox{$>$}}}}

\newcommand{\beq}{\begin{equation}}
\newcommand{\eeq}{\end{equation}}
\def\beqa{\begin{eqnarray}}
\def\eeqa{\end{eqnarray}}
\def\hMpc{$h^{-1}\,{\rm Mpc}$}
\def\hkpc{$h^{-1}\,{\rm kpc}$}

\def\head{
 \vbox to 0pt{\vss
                   \hbox to 0pt{\hskip 440pt\rm LA-UR-10-07069\hss}
                  \vskip 25pt}}

\title[Exploring the nature of three MW merger analogues]
{A high fidelity Milky Way simulation with Kraken, Gaia-Enceladus, and Sequoia analogues: clues to their accretion histories}
\author[Garc\'{i}a-Bethencourt et al.] 
      {Guacimara Garc\'{i}a-Bethencourt$^{1,2}$ \thanks{E-mail: \href{alu0100971361@ull.edu.es}{alu0100971361@ull.edu.es}}, Chris B. Brook$^{1,2}$, Robert J. J. Grand$^{1,2,3}$, 
      \newauthor
      Daisuke Kawata$^{4}$\\
$^{1}$Universidad de La Laguna. Avda. Astrof\'{i}sico Fco. S\'{a}nchez, La Laguna, Tenerife, Spain\\
$^{2}$Instituto de Astrof\'{i}sica de Canarias, Calle Via L\'{a}ctea s/n, E-38206 La Laguna, Tenerife, Spain\\
$^{3}$Astrophysics Research Institute, Liverpool John Moores University, 146 Brownlow Hill, Liverpool, L3 5RF, UK\\
$^{4}$Mullard Space Science Laboratory, University College London, Holmbury St. Mary, Dorking, Surrey, RH5 6NT, UK\\
}

\begin{document}

\date{Accepted xxxx. Received xxxx; in original form xxxx}

\pagerange{\pageref{firstpage}--\pageref{lastpage}} \pubyear{2023} 

\maketitle

\label{firstpage}

\begin{abstract}
Within a simulated Milky Way-like galaxy, we identify and analyse analogues of the Gaia-Enceladus (GE), Kraken and Sequoia mergers that each matches remarkably well observational results, including in velocity and chemical abundance space, and their distributions in the  \textit{j$_z$}-Energy plane. The Kraken analogue is the earliest merger and has the highest total mass ratio. Consistent with previous studies, it is chemically indistinguishable from old in-situ stars at the time of its accretion. The GE and Sequoia analogue  events accrete at similar times in our simulation, both along filaments but from opposite sides of the main galaxy. The mean stellar ages of the GE and Sequoia analogues are both similar and, from our simulation results, we see that they can be separate entities and still naturally reproduce the observed properties of their stellar remnants at the present day, including the significant retrograde velocities of the Sequoia analogue remnant stars and the difference in the tracks of the two galaxies through chemical abundance space.  Our results provide supporting information about the properties of these three merger events, and show for the first time that they can all be reproduced with a fully cosmological simulation, providing a possible self consistent evolutionary pathway for the Milky Way's formation.  
\end{abstract}

\noindent
\begin{keywords}
galaxies: formation -- galaxies: evolution -- galaxies: kinematics and dynamics -- galaxies: abundances -- methods: numerical
\end{keywords}

\section{Introduction} \label{sec:intro}

The formation and evolution of the Milky Way and its merger history is an active field of study, with detailed information provided in particular by the \textit{Gaia} satellite \cite[]{gaiacollaboration2016, gaiacollaboration2018}. There is evidence that mergers have had an important impact on the development and growth of our galaxy \citep[e.g.][]{Helmi_2018}. Merged satellite galaxies have left imprints of their accretion, especially in the stellar halo (e.g. \citealt{matsuno2019, helmi2020}), meaning that details of mergers can be studied at the present day through their kinematics and/or chemical abundances, properties that are likely to remain nearly constant for long periods of time \citep[e.g.][]{johnston1996, johnston1998,helmi_white1999}. Analysis of chemical abundance ratios also provides information of the environment in which accreted stars were born, and hence, about the properties of their parent satellite galaxy \citep[e.g.][]{freeman_hawthorn2002}.

Several satellite accretion events have been identified in the Milky Way. For instance, the  Sagittarius (Sgr) dwarf galaxy \citep{ibata1994}, with a first passage into the Milky Way around 6 Grys ago \citep{ruizlara2020}, has left a stream of stars \citep{ibata2020, vasiliev2020} in an ongoing accretion process.

There is evidence of earlier merger events in the Milky Way. Gaia-Enceladus (GE; \citealt{Helmi_2018}) or Gaia-Sausage \citep{belokurov2018} was accreted $\sim$10 Gyrs ago (see also \citealt{gallart2019birth, chaplin2020}) with an initial stellar mass of $5 \times 10^{8}$ - $5 \times 10^{9}$~M$_{\odot}$ (see also \citealt{mackereth2019, vincenzo2019}). This is believed to be the last significant merger of our galaxy and was first identified  as a stellar population with large radial motion, highly eccentric orbits, and low metallicity \citep{Chiba_2000,Brook_2003}. GE dominates the region within the inner stellar halo, and a significant amount of its stellar debris is found in the solar neighbourhood, with a slight retrograde motion (\citealt{villalobos2008}). GE stars have mean metallicities of [Fe/H]$\sim$$-1.2$ to $-1.5$ (e.g. \citealt{nissen_schuster2011}) and low [$\alpha$/Fe] values (e.g. \citealt{meza2005,nissen_schuster2011, haywood2018, mackereth2019}).

There is also evidence that the accretion of GE had a significant impact on the primitive Milky Way, when a fraction of stars in the primordial disc were heated (e.g. \citealt{Zolotov_2009, gallart2019birth}) to more halo-like kinematics. This event, known as the ‘splash’ \citep{belokurov2020}, created an in-situ population of ‘hot thick disc’ \citep{Helmi_2018, dimatteo2019} stars.

An even more ancient merger, dubbed Kraken \citep{kruijssen2020, massari2019}, has been proposed to explain  the age-metallicity relation of globular clusters in the Milky Way. This early satellite has been estimated to have M$_{\star}\sim 2 \times 10^{8}$~M$_{\odot}$ at the time of infall, about 11 Gyrs ago \citep{kruijssen2020}; less massive than GE. However, it might have been of great importance in the evolution of the galaxy, due to a larger mass ratio between Kraken and the central galaxy at the time of its accretion. The remnant stars from Kraken are thought to have high binding energies, residing in the most inner parts of the Galaxy \citep{horta2021}. The properties and even existence of Kraken are still quite uncertain, because its early accretion means that the stars have become mixed in phase space and cannot be easily identified using kinematics alone. Moreover, it has been suggested by \cite{orkney2022} that the Kraken has similar chemical abundances than that of the main galaxy at the time of infall, and hence, it might also be difficult to identify through the analysis of chemical abundances. Conversely, a recent study by \cite{horta2022} advocates chemical distinction between Kraken/Heracles \citep{horta2021} and the low-metallicity in-situ population of stars known as \textit{Aurora} \citep{belokurov_kravtsov2022}.

A smaller and later merger structure has been named Sequoia \citep{myeong2019}, with an estimated stellar mass of $\sim 5 \times 10^{7}$~M$_{\odot}$ and accretion time around 9 Gyrs ago. Its stellar debris has mainly retrograde orbits and low binding energies, and is suggested to contribute to the retrograde motion found in the outer galactic halo, along with GE \citep{helmi2017, koppelman2019}. Some debate has occurred regarding chemical abundances of Sequoia \citep{matsuno2019, monty2020, naidu2020} and whether it consisted of a single population, and whether it overlaps GE stars. A recent study \citep{matsuno2022} aimed to clear up this picture by using high-precision abundances, and found a clear separation between Sequoia and GE stars, with lower metallicity ([Fe/H]) and lower alpha element abundances, ([$\alpha$/Fe]).

In this paper we identify analogues of GE, Sequoia, and Kraken within a simulated Milky Way analogue galaxy. This Milky Way analogue has been shown to match a range of our Galaxy's observed properties, including detailed abundance patterns within the thick and thin discs, \citep{brook2020chemical}, the vertical and radial metallicity gradients \citep{miranda2016} of these populations, and the accreted and in-situ halo population related to the ‘splash’ caused by the GE merger \citep{gallart2019birth}. In this study we explore the features of these three mergers in detail, including phase space velocities, energy-angular momentum, and chemical abundances, and find excellent agreement with the observed accreted satellites. This allows us to explore details of the mergers to provide hints on their sequence and information of their infall.

This paper is structured as follows: in Section~\ref{sec:sim} we present details of the simulation used for the study. In Section~\ref{sec:res} we  specify the main properties of each merger analogue, such as masses and accretion redshifts, kinematic phase space, chemical abundances, spatial distribution, star formation histories, and infall details. Next, Section~\ref{sec:disc} holds the discussion of these results, and in Section~\ref{sec:concl} we conclude and summarise the analysis and outcomes.

\section{Simulation} \label{sec:sim}

\subsection{The MaGICC simulations}\label{sec:sim-1}
The simulated galaxy of analysis is part of the MaGICC project \citep{stinson2012, brook2012}, a set of cosmological hydrodynamical simulations that reproduce observed galaxy scaling relations \citep{brook2012}. These simulations are  evolved with the Smoothed Particle Hydrodynamics (SPH; \citealt{gingold_monaghan1977}) code \textsc{gasoline} \citep{wadsley2004}. 

\begin{table}
	\centering
	\begin{tabular}{lccc} 
		\hline
		 & Kraken & Gaia-Enceladus & Sequoia \\
		\hline
		\textit{z}$_{\rm \star max}$, \textit{z}$_{\rm \star merge}$ & 5.53, 4.79 & 2.87, 2.65 & 2.13, 1.88 \\
		\textit{z}$_{\rm mixed}$ & $\sim$ 2.87 &  $\sim$ 1.57 & $\sim$ 0.82 \\ 
		M$_{\rm total}$ [M$_{\odot}$] & 1.26 $\times 10^{10}$ & 5.70 $\times 10^{10}$ & 6.17 $\times 10^{9}$* \\
		M$_{\rm stars}$ [M$_{\odot}$]  & 2.57 $\times 10^{7}$ & 3.55 $\times 10^{8}$ & 2.64 $\times 10^{7}$ \\
		M$_{\rm gas}$ [M$_{\odot}$] & 1.88 $\times 10^{9}$ & 1.11 $\times 10^{10}$ & 1.11 $\times 10^{9}$* \\
		M$_{\rm t, merger}$/M$_{\rm t, main}$ & 0.306 & 0.145 & 0.009* \\
		\hline
	\end{tabular}
	\caption{Main properties of the sample of merger analogues. \textit{z}$_{\rm \star max}$ and \textit{z}$_{\rm \star merge}$ define the redshift before the accretion and the redshift where the merger begins, i.e. there is a close encounter with the  central galaxy and significant disruption. Specifically, \textit{z}$_{\rm \star max}$ also defines the redshift at which the stellar mass of each merger is maximum. \textit{z}$_{\rm mixed}$ is an approximate redshift at which the satellite stars have become disrupted and are spatially spread throughout the host galaxy's halo, and thus not in a spatially coherent configuration. The last row indicates the total mass ratio between the mergers and the main galaxy at \textit{z}$_{\rm \star max}$. All masses are calculated at \textit{z}$_{\rm \star max}$, except those marked with an asterisk (*), which are measured at the time of peak total mass of sim$\_$Sequoia ($z=2.28$) so as to consider possible uncertainties due to tidal and ram pressure stripping during the merger's pericentric passage.}
	\label{tab:merger_properties}
\end{table}

Gas cooling is due to H, He and several metal lines \citep{shen2010}, including Compton and radiative cooling.  Metal cooling uses \textsc{cloudy} \citep{Ferland_1998}, and assumes an uniform UV ionising radiation background \citep{haardt_madau1996}.  The  diffusion of metals (e.g. C, O, Fe, Si, N, Mg) is treated by using a turbulence mixing algorithm \citep{wadsley2008}.

Star formation processes take place in high-density regions, when gas reaches a temperature of $T < 10^{4}$~K and a density of $n_{th} < 9.3 $~cm$^{-3}$. The Kennicutt-Schmidt law \citep{schmidt1959, Kennicutt_Jr__1998} handles the amount of gas that is turned into stars:

\begin{equation}
    \frac{\Delta M_{\star}}{\Delta t} = c_{\star} \frac{M_{gas}}{t_{dyn}}
    \label{eq: kennicutt_schmidt_law}
\end{equation}

\noindent where $\Delta M_{\star}$ is the mass of each star particle formed, $\Delta t$ is the timestep between star formation events, $M_{gas}$ is the mass of the gas particles, $c_{\star}$ is the star formation efficiency, and $t_{dyn}$ is the dynamical or free-fall time of the gas particles. The fraction of gas that is converted to stars during $t_{dyn}$ is given by $c_{\star} = 0.167$. The star formation rate is $\rho_{gas}^{1.5}$, being $\rho_{gas}$ the gas density.

The star particles formed are collisionless and represent clusters of co-eval stellar populations, with a Chabrier \citep{Chabrier_2003} Initial Mass Function (IMF). 

Stellar feedback helps to regulate star formation in the system and is implemented via two different types. The first type refers to the Early Stellar Feedback (ESF; \citealt{stinson2012}), that accounts for the energetic radiation ejected by young massive stars before reaching the supernova phase. Also included is feedback due to type Ia supernovae (SNIa) and type II supernovae (SNII), that introduce energy and eject metals into the interstellar medium (ISM). The energy from supernovae is implemented via the blast-wave formalism \citep{stinson2006}. Metals are ejected from both types of supernovae and via the stellar winds driven from Asymptotic Giant Branch (AGB) stars. This mass is distributed to the nearest neighbour gas particles through the smoothing kernel \citep{stinson2006}.

Our Milky Way analogue is labelled {\fontfamily{pcr}\selectfont g15784} from the MUGS set of galaxies \citep{stinson2010}. At $z = 0$, there are $\sim$ 4 million particles within the virial radius of R$_{vir} = 242$~kpc. The average mass of star, gas, and dark matter particles are: $M_{\rm star, p} = 3.85 \times 10^{4}$~M$_{\odot}$, $M_{\rm gas, p} = 2.05 \times 10^{5}$~M$_{\odot}$, $M_{\rm DM, p} = 1.11 \times 10^{6}$~M$_{\odot}$, respectively. The total, stellar and gas mass of the simulated galaxy (measured within R$_{vir}$) are the following: $M_{\rm total} = 1.5 \times 10^{12}$~M$_{\odot}$, $M_{\star} = 8.28 \times 10^{10}$~M$_{\odot}$, $M_{\rm gas} = 13.50 \times 10^{10}$~M$_{\odot}$, respectively.

We use the \textsc{amiga} Halo Finder (AHF; \citealt{gill2004, knollmann2009}) to identify and trace halos. This process has uncertainties such as the frequency of the simulation outputs, which is $\sim0.214$ Gyrs between snapshots, and it needs to be considered when interpreting the accretion redshifts of each merger.

The analysis is partly performed with the \textsc{pynbody}\footnote{\urlstyle{rm}\url{https://pynbody.github.io/pynbody/}} package \citep{pynbody}.

\section{Results}\label{sec:res}

\begin{table}
	\centering
	\begin{tabular}{lccc} 
		\hline
		 & Inner bulge & Outer bulge & Solar region \\
		 & (0-2 kpc)  & (2-3 kpc)  & (7-9 kpc)  \\
		\hline 
		Kraken & 54\% / 0.02\% & 14\% / 0.05\% & 2\% / 0.01\% \\
		Gaia-Enceladus & 7\% / 0.05\% & 7\% / 0.40\% & 11\% / 0.66\% \\
		Sequoia & - & - & 1\% / 0.01\% \\
		\hline
	\end{tabular}
	\caption{Percentage of stars from the various merged  galaxies that end in the inner and outer bulge, and the solar region, and the percentage of stars in those regions that come from those merged galaxies.} 
	\label{tab:contribution}
\end{table}

To obtain the stellar distributions and properties of each merger, we trace merged satellite galaxies since $z = 5.53$, the earliest snapshot,  following their evolution and disruption, when their stars become part of the central galaxy. 

We identified three mergers that are analogues to proposed accreted structures in the Milky Way, namely: Kraken, Gaia-Enceladus, and Sequoia. The main properties of these analogue galaxies are shown in Table~\ref{tab:merger_properties}, and we will refer to them as sim$\_$Kraken, sim$\_$Enceladus and sim$\_$Sequoia. The total, stellar and gas masses are measured prior to the accretion, in the timestep at which the stellar mass of the mergers is maximum, \textit{z}$_{\rm \star max}$. The total mass ratio at \textit{z}$_{\rm \star max}$ between the merging satellite and the central galaxy is also shown in the table: sim$\_$Kraken has the highest merger ratio, whilst  sim$\_$Enceladus is the most massive accreted satellite. 

For further detail, \textit{z}$_{\rm \star merge}$ represents the redshift at which each merger can no longer be identified by the AHF, whilst at \textit{z}$_{\rm \star max}$ they are still separate halos. Additionally, at \textit{z}$_{\rm \star merge}$ the stellar mass of the merger analogues has decreased due to stellar evolution and/or tidal stripping by the close interaction with the host galaxy.

In Table~\ref{tab:merger_properties} we also include the approximate redshifts when the satellite analogues are spatially mixed within the host halo, i.e. their stars begin to form a spheroidal-like structure. Hence, this redshift, defined as \textit{z}$_{\rm mixed}$, is estimated visually from the stellar distribution of the mergers in coordinate space. It can be seen that this process requires a significant amount of time.

\begin{figure}  
\includegraphics[width=3.5in,height=4.0in, trim={1.0cm 0.5cm 0.5cm 0cm}, clip]{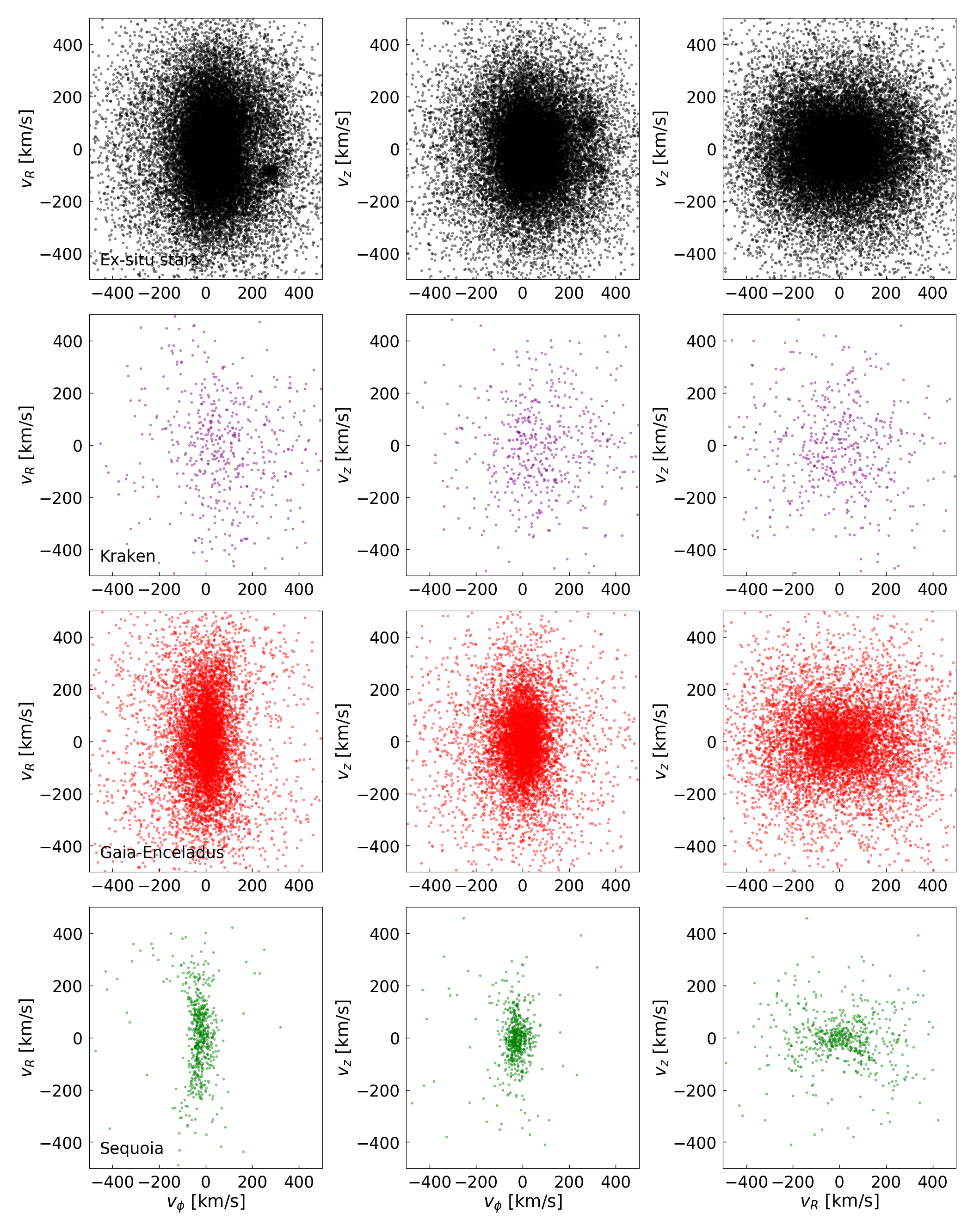}
\caption{Distributions of stars in the three-dimensional velocity space at $z = 0$. From top to bottom, we show the distributions of all ex-situ stars (in black), sim$\_$Kraken (in purple), sim$\_$Enceladus (in red), and sim$\_$Sequoia (in green), respectively. }
\label{fig:velocities_panel} 
\end{figure}

\subsection{Kinematics and Integrals of Motion (IoM)}\label{sec:res-1}

The velocity distributions at $z = 0$ in cylindrical coordinates are shown in Fig.~\ref{fig:velocities_panel}. The left column represents the radial velocity in the plane (V$_R$) of the disc versus the rotational velocity (V$_\phi$), the central column shows the velocity perpendicular to the disc (V$_Z$) vs V$_\phi$, and the last column shows V$_Z$ versus V$_R$. The sim$\_$Kraken distribution has prograde rotation ($<v_{\phi}> =$ 74~km~s$^{-1}$), and large dispersion in all three velocity components. The sim$\_$Enceladus shows the characteristically large radial velocities, and a small mean retrograde rotation of $-5$~km~s$^{-1}$. For sim$\_$Sequoia, we also see a large range of V$_R$, and more significant retrograde rotation ($<v_{\phi}> = -33$~km~s$^{-1}$).

\begin{figure}  
\includegraphics[width=3.38in,height=5.3in]{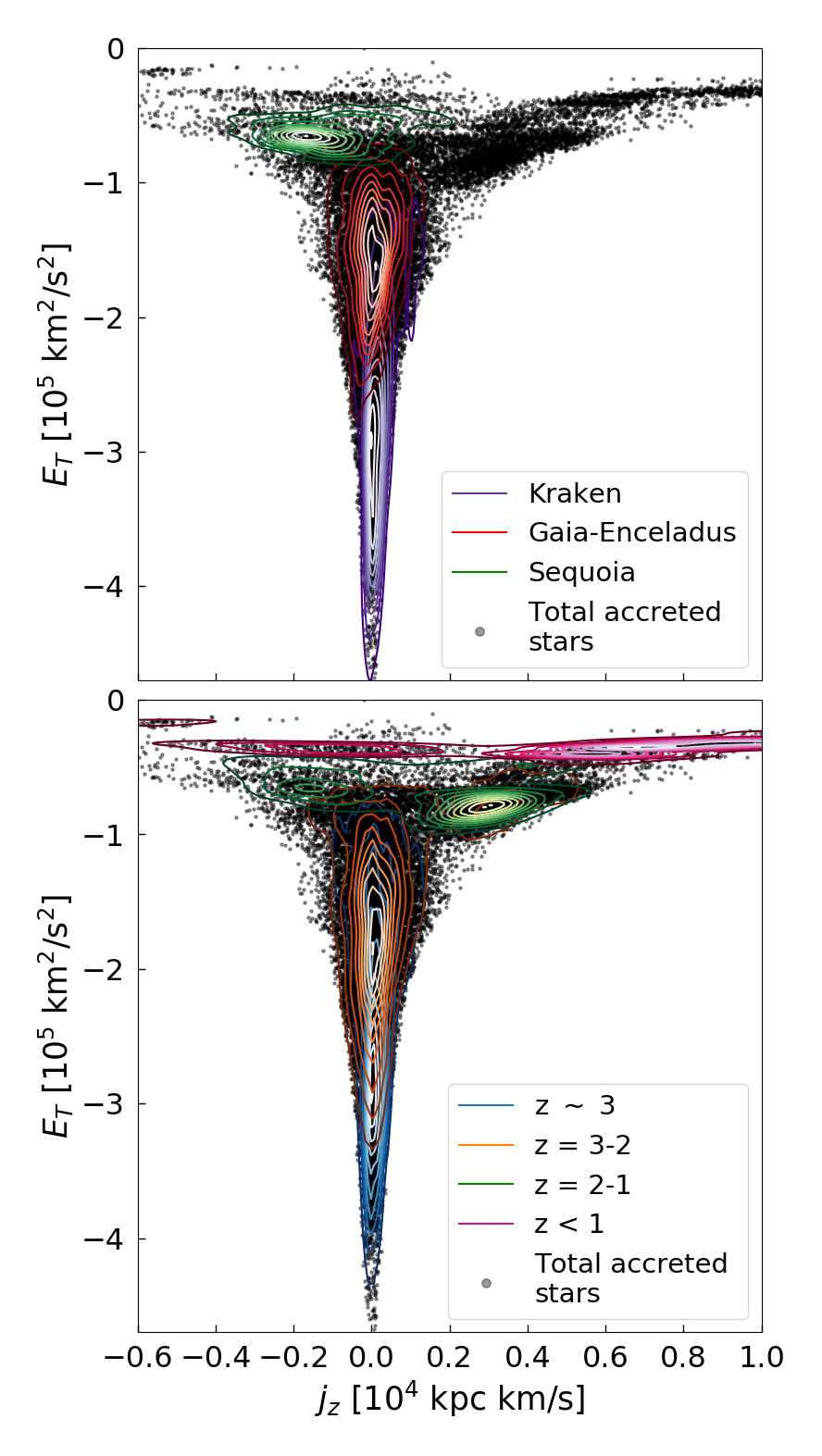}
\caption{Distribution of accreted stars in IoM space at $z = 0$. We show the space defined by the total specific energy and z-component of the angular momentum. On the top panel, we present all ex-situ stars (black), sim$\_$Kraken (purple), sim$\_$Enceladus (red), and sim$\_$Sequoia (green). On the bottom panel, we show all ex-situ stars (black), stars accreted at $z \sim 3$ (blue), stars accreted at $z = 3-2$ (orange), stars accreted at $z = 2-1$ (green), and stars accreted at $z < 1$ (pink). Contours are normalised to the maximum density value of each distribution and shown as fractions from 0.1 to 1.0.}
\label{fig:IoM} 
\end{figure}

On the top panel of Fig.~\ref{fig:IoM}, we show the stellar distributions in the space defined by the total specific energy vs the specific z-component of the angular momentum, where each colour represents a different merger. The total energy is normalised to the maximum energy of all stars belonging to the main galaxy, so the limit of energetically bound stars is set to zero. We can see that sim$\_$Kraken (in purple) has low energy and low angular momentum, with mean values of ($E_T$, $j_z$) $=$ ($-$2.6 $\times 10^{5}$~km$^{2}$~s$^{-2}$, $1.7 \times 10^{2}$~kpc~km~s$^{-1}$). This is a result of the early accretion of the original satellite galaxy, as it was already close to the bottom of the potential well of the system at that time. There is scattering to higher energies, but all values are below $-1.0 \times 10^{5}$~km$^{2}$~s$^{-2}$. The distribution of sim$\_$Enceladus stars (in red) is found at ($E_T$, $j_z$) $=$ ($-1.5 \times 10^{5}$~km$^{2}$~s$^{-2}$, $2.7 \times 10^{1}$~kpc~km~s$^{-1}$) in average. The region defined by this merger agrees with results from studies of observed GE stars \citep{massari2019,koppelman2019,ruiz-lara2022b}. Lastly, sim$\_$Sequoia has low binding energies and a relatively small dispersion. It also shows negative angular momentum, reflecting its retrograde rotational velocity. The mean values of energy and angular momentum of this merger are the following: ($E_T$, $j_z$) $=$ ($-6.4 \times 10^{4}$~km$^{2}$~s$^{-2}$, $-9.9 \times 10^{2}$~kpc~km~s$^{-1}$). These results are also consistent with the area populated by observed Sequoia stars in previous analysis \cite[e.g.][]{massari2019,wang2022}. 

Compared to Fig.~\ref{fig:velocities_panel}, the identification of all three mergers is clearer in the space of the Integrals of Motion, since these are better conserved quantities. 

The bottom panel of Fig.~\ref{fig:IoM} shows the distributions of accreted stars at different redshift intervals. It clearly shows how accretions at different redshifts occupy different regions of the energy space.  As well as the three accreted satellites focussed on in this study, there is also a prograde structure of stars coloured in green. Although it is prominent in this plot, this merger is later and has lower mass than GE, with a stellar mass of 7.62 $\times 10^{7}$~M$_{\odot}$ at $z_{\rm \star max} = 1.76$.

We find that the sim$\_$Enceladus event has an impact on previous accreted stellar structures in terms of their kinematics at the time of its infall (between $z = 2.87$ and $z = 2.65$). This effect is seen as an increase in the dispersion of the vertical velocity component (V$_Z$) of early accreted distributions. The implication is that the merger of sim$\_$Enceladus ‘splashed’ ex-situ stars, as well as splashing in-situ forming (thick) disc stars into the halo \citep[see also][]{grand2020}.

\subsection{Contribution of accreted stars to the galactic bulge and the solar region}\label{sec:res-2} 

We next study the contribution of accreted stars to the galactic bulge and to the solar neighbourhood at the present day. We define the inner bulge as a sphere with a radius of 2 kpc, while the outer bulge is defined as a concentric spherical layer between 2 and 3 kpc from the galactic centre. The solar region is defined as a spherical shell between 7 and 9 kpc. The results are displayed in Table~\ref{tab:contribution}, where each row contains the percentage of stars from each progenitor satellite that are found within the selected areas, along with the percentage that those stars represent with respect to the total stellar amount of each region. 

\begin{figure}  
\includegraphics[width=3.3in,height=2.2in]{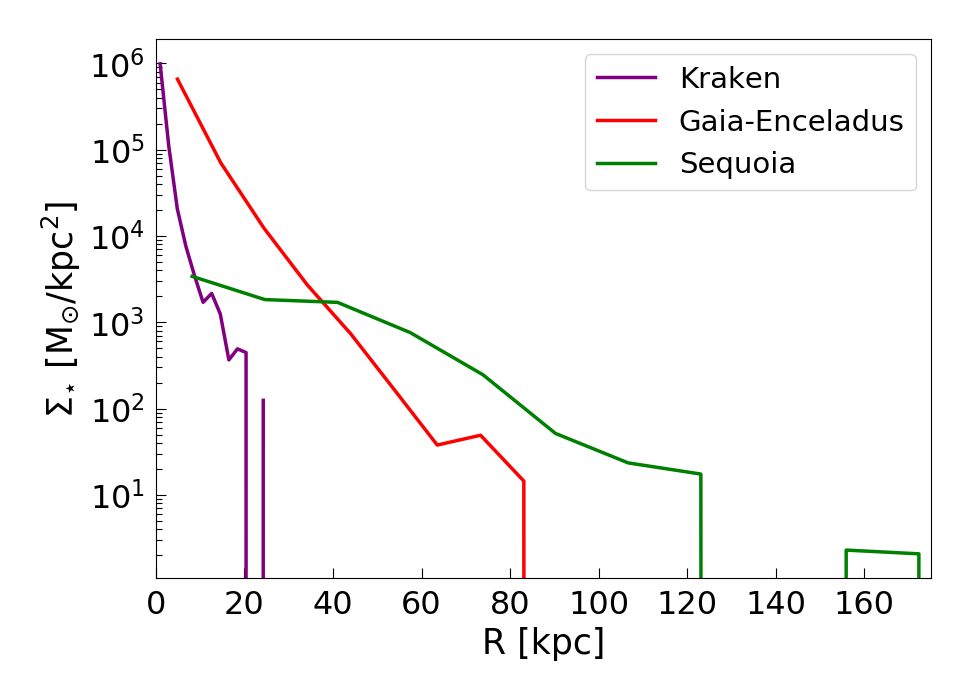}
\caption{Surface density profiles of stars from the merger analogues with respect to the distance from the centre of the galaxy at $z = 0$.} 
\label{fig:density_profiles} 
\end{figure}

More than half of the progenitor satellite of sim$\_$Kraken is found in the inner 2 kpc of the simulated galaxy, whilst $\sim$14 per cent and $\sim$2 per cent of it is located in the outer bulge and solar region, respectively. However, the contribution of this merger to the present-day stellar populations is small at all radii, ranging in 0.01-0.05 per cent. 

The later,  more massive merger of sim$\_$Enceladus is more extended in radial distance, and making larger contribution to the solar neighbourhood, having $\sim$11 per cent of its stars residing there. The contribution of sim$\_$Enceladus to the innermost regions is relatively small. As for sim$\_$Sequoia, it is even more radially extended, barely contributing to any of the selected areas, with only a $\sim$1 per cent of the original satellite within the solar region. These results reflect that later mergers with higher angular momentum contribute stars in more outer parts of the galaxy and are less bound (see also Fig.~\ref{fig:IoM}).

\subsection{Stellar density profiles}\label{sec:res-3}

In Fig.~\ref{fig:density_profiles} we present the surface density profiles of the stars for the merger analogues at the present time. It is seen that sim$\_$Kraken and sim$\_$Enceladus have densities above $10^{5}$~M$_{\odot}$~kpc$^{-2}$ within 20 kpc from the centre of the galaxy, while the density of sim$\_$Sequoia  is under $10^{4}$~M$_{\odot}$~kpc$^{-2}$. This is consistent with the results presented in Table~\ref{tab:contribution}, and shows that sim$\_$Sequoia is more radially dispersed within the galaxy, and hence, it is more likely to contribute more to the outer regions, i.e. above 20 kpc. The stellar debris from sim$\_$Kraken contributes to the oldest population of stars in the bulge (see Table~\ref{tab:contribution}), sim$\_$Enceladus contributes to the bulge and inner halo,  whilst sim$\_$Sequoia tends to provide  stars to the outer halo.

\subsection{Star Formation Histories}\label{sec:res-5}

\begin{figure}  
\includegraphics[width=3.35in, height=5.5in]{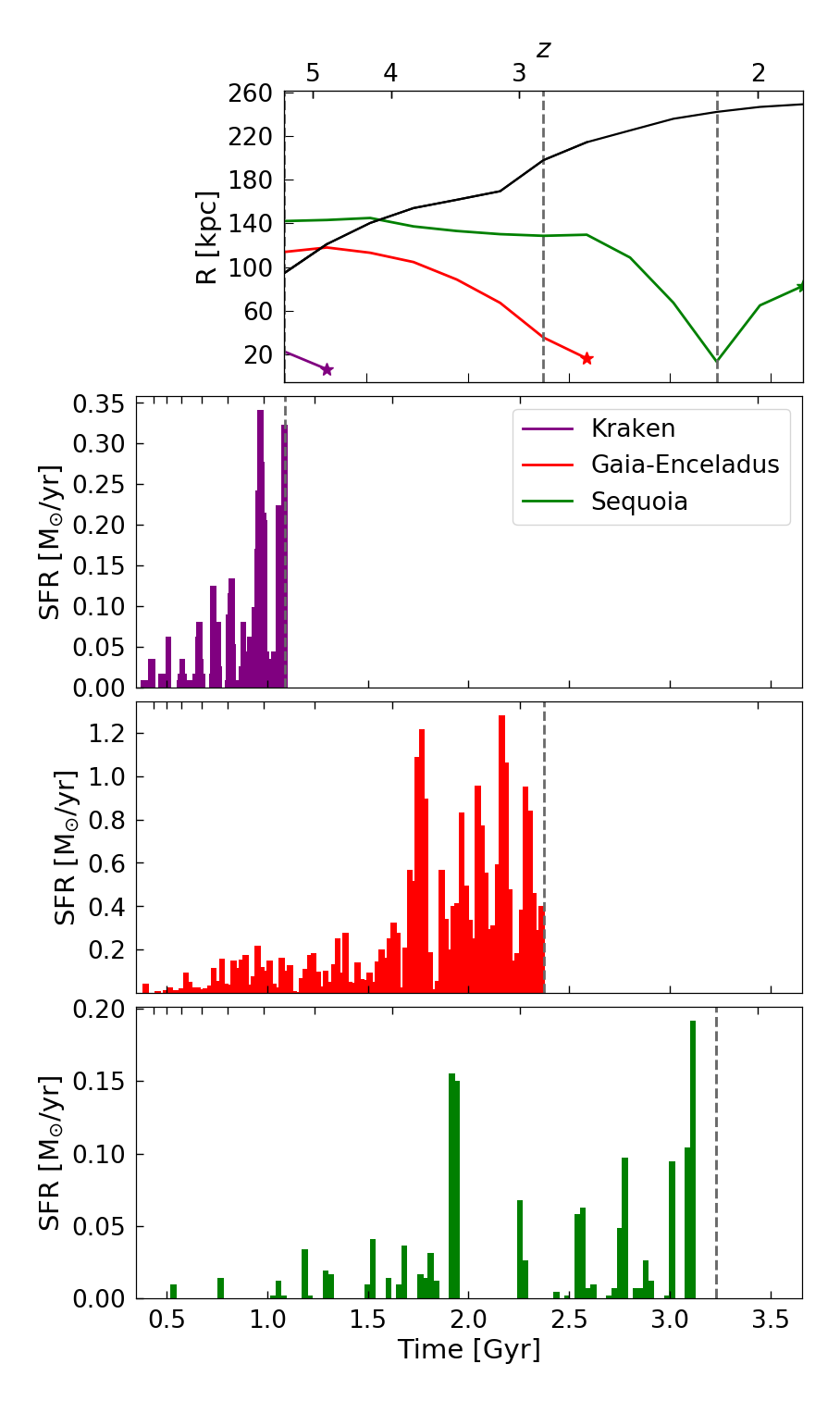}
\caption{On the top panel, temporal evolution of the mean radial distance of stars belonging to sim$\_$Kraken (purple), sim$\_$Enceladus (red) and sim$\_$Sequoia (green). The evolution of the virial radius of the galaxy is shown in black. The dashed vertical lines specify the redshift of maximum stellar mass prior to the accretion into the main system, this is at \textit{z}$_{\rm \star max}$. The star-shaped marks indicate the moment of the merger, i.e. \textit{z}$_{\rm \star merge}$. The lower panels show the star formation histories of the mergers measured at $z = 0$ and using a number of 100 bins.}
\label{fig:sfh_large} 
\end{figure}

The top panel of Fig.~\ref{fig:sfh_large} shows the average radial distance of the mergers from the galactic centre at each time. We indicate the positions of \textit{z}$_{\rm \star max}$ and \textit{z}$_{\rm \star merge}$ with vertical dashed lines and star-shaped marks, respectively. We also indicate the growth of the galaxy's viral radius in black. At the first output timestep recorded at redshift $\sim$5.5, sim$\_$Kraken is already only 20 kpc from the central galaxy. 

We also show the star formation histories of the three mergers in Fig.~\ref{fig:sfh_large}. The three lower panels display increasing star formation rates (SFRs) in a timespan of $\sim$3 Gyrs. The star formation history of these mergers suddenly stops after several peaks of high star formation activity, which could be caused by a compression of their gas during the process of accretion, as seen in \cite{dicintio2021}.  The cease in star formation coincides with the time of infall of sim$\_$Kraken and sim$\_$Enceladus, as indicated with the dashed lines in the panels. On the contrary, sim$\_$Sequoia has its star formation halted less than 0.5 Gyrs before the accretion into the galaxy, that may indicate a loss of the gas reservoir by tidal or ram pressure stripping at that epoch \citep{simpson2018}. 

The average age of stars in the debris of the satellites is not only driven by infall time, but also by the SFR, with an increasing star formation rate causing the average age to be closer to the infall time whilst a steady SFR would mean the average age of stars is earlier. Thus, although sim$\_$Sequoia has a later infall time than sim$\_$Enceladus in this simulation, the average age of its stars are quite similar. Another effect that needs to be considered is that ongoing star formation may occur during the accretion of sim$\_$Enceladus, which can also affect the calculation of accretion time from observed stellar ages \citep{dicintio2021}.

\subsection{Chemical abundances: the [O/Fe] vs [Fe/H] plane}\label{sec:res-6}
\begin{figure}  
\includegraphics[width=3.3in,height=1.3in, trim={1.0cm 0.5cm 0.5cm 0.5cm}, clip]{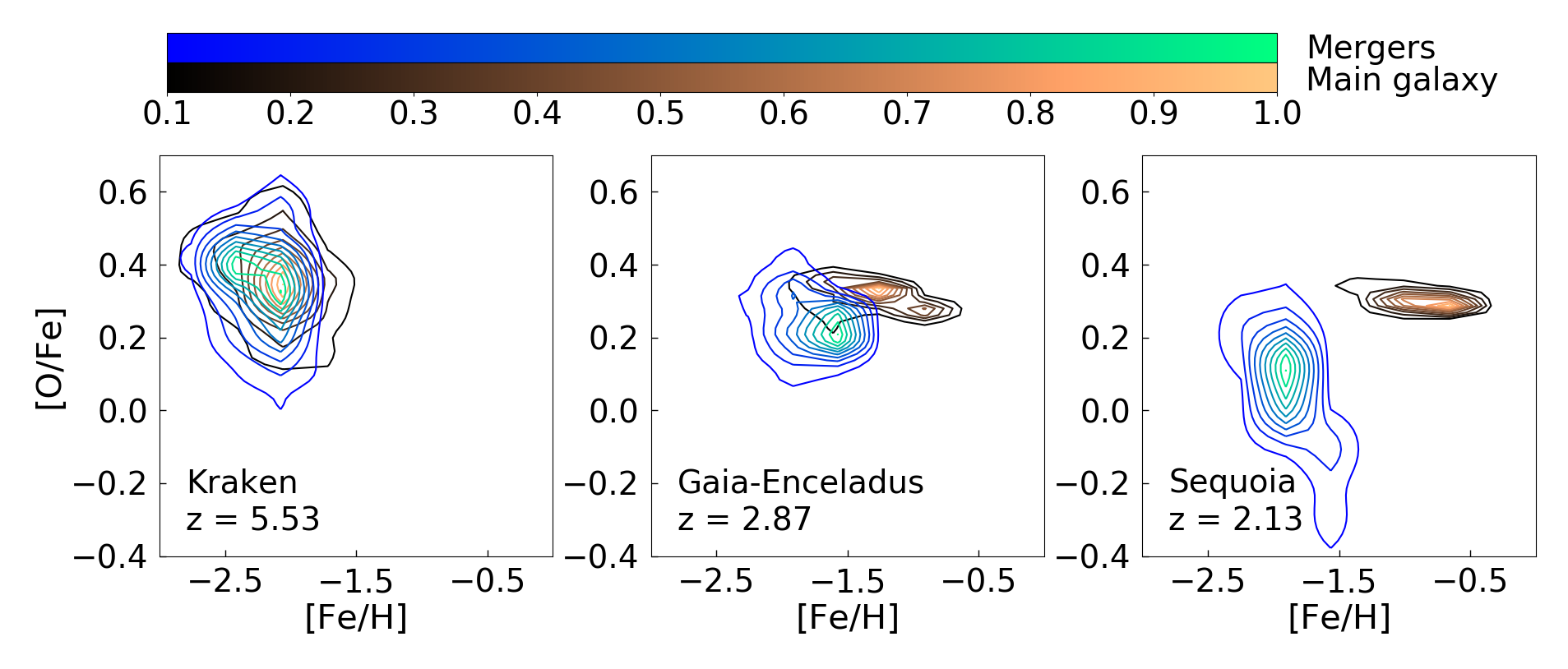}
\caption{[O/Fe] versus [Fe/H] of stars from the main galaxy and from the mergers at their respective times of infall, \textit{z}$_{\rm \star max}$. From left to right, we show the sim$\_$Kraken, sim$\_$Enceladus and sim$\_$Sequoia distributions in blue-green contours. The black-orange background represents the stellar distribution of the main galaxy at each time. Contours are normalised to the maximum density value of each distribution individually, and shown from 0.1 to 1.0.}
\label{fig:feh_ofe_large} 
\end{figure}

We explore the metallicities of the stars from our selection of merger analogues in the [O/Fe] versus [Fe/H] plane, as shown in Fig.~\ref{fig:feh_ofe_large}. These chemical abundances are calculated prior to the accretion of each satellite galaxy, at \textit{z}$_{\rm \star max}$. Merged stars are shown as distributions of blue-green scale of contours, whilst the black-orange background represents the stars belonging to the main galaxy at each time. All contours are normalised so that they display the levels from 0.1 to 1.0 times their maximum density value. 

We can see the distribution associated to sim$\_$Kraken and the main galaxy in the first panel of the figure. These are placed at low metallicities, with mean [Fe/H] of $-2.77$ and $-2.49$ dex, respectively. Both distributions show a high level of overlap, having little differences between them as discussed in \cite{orkney2022} for the [Mg/Fe] vs [Fe/H] plane. 

In the case of sim$\_$Enceladus, the difference with the main progenitor in the [Fe/H] = [O/Fe] plain is clear, with just a small overlap, as is observed \citep[e.g.][]{matsuno2022}. As previously noted in \cite{brook2020chemical}, the sim$\_$Enceladus is accreted around half a Gyr earlier than the estimated time when the real GE was accreted into the Milky Way, meaning it is accreted with lower metallicity. The offset from the main progenitor in [O/Fe]-[Fe/H] space is nevertheless clear in both the simulations and observations, which has been attributed to different star formation efficiencies of the main progenitor and the accreted galaxy \citep{brook2020chemical}.

For the last panel, we see that sim$\_$Sequoia follows a different track in [Fe/H] = [O/Fe] space again, reflective of its lower mass and even lower star formation efficiency. The region covered by our sim$\_$Sequoia distribution is in good agreement with the results presented in \cite{matsuno2022} (see e.g. fig. 10). Here, they find a clear abundance distinction between Sequoia and GE of $\sim$0.1-0.2 dex, which is also consistent with our results.

As seen in Fig.~\ref{fig:feh_ofe_large}, we also expect that these merger remnants follow different age-metallicity sequences as seen in \cite{massari2019} and \cite{kruijssen2020}.

\subsection{Accretion along filamentary structure}\label{sec:res-7}

In Fig.~\ref{fig:filaments} we show the two projections of the position of sim$\_$Enceladus (red) and sim$\_$Sequoia (green) stars at the time prior to the accretion of sim$\_$Enceladus. On the top panel, the central galaxy is aligned face on, whilst in the bottom panel the central galaxy is aligned edge on. The background shows the gas density as shown on the color bar on the right.  The two satellites, sim$\_$Enceladus and sim$\_$Sequoia  are both accreted from the most prominent filament of that epoch, but from opposite directions. This provides a possible explanation as to their similarly radial dominated orbits. 

We note that, unlike sim$\_$Enceladus and sim$\_$Sequoia, which infall down the dominant filament in retrograde orbits, the prograde structure found in Fig.~\ref{fig:IoM} is accreted along a less dense and less prominent filament.

\section{Discussion}\label{sec:disc}

In our simulation, sim$\_$Kraken merges at a very early time, ending deep inside the potential well with its stars predominately in the bulge according to the results presented in Table~\ref{tab:contribution}. As in other studies \citep[e.g.][]{kruijssen2020, orkney2022}, our Kraken analogue is the most significant merger in terms of having the highest mass ratio. The early merger  and short formation time prior is reflected in the high $\alpha$ abundances \cite[]{naidu2022}. As also shown in \cite{orkney2022}, the failure to reach a downturn (or ‘knee’) in the [$\alpha$/Fe]-[Fe/H] plane means that it is not possible to distinguish the abundances of Kraken from other stars forming in the central galaxy at the same time. Nonetheless, it has to be considered that there may be certain differences from in-situ stars in other non-alpha elements \citep[e.g.][]{horta2021}.

Particularly, this Kraken analogue is $\sim 10$ times less massive in stars than that found in \cite{kruijssen2020}, and $\sim 100$ times less massive than the Kraken-like merger studied in \cite{orkney2022}. Since the mass growth is large for early mergers, the accreted mass is strongly dependent on the redshift of accretion. However, it is important to consider that our sim$\_$Kraken still matches the high $\alpha$ abundance distribution, together with the main progenitor, during this phase of rapid mass growth.

The accretion of sim$\_$Enceladus in this simulation  has been previously studied \cite[]{gallart2019birth,brook2020chemical}. The accreted stars of sim$\_$Enceladus resemble those of the observed GE in chemo-dynamical space, including the ‘sausage’ shape in the radial-rotational velocity plane \cite[]{belokurov2018} and downturn in the [$\alpha$/Fe]-[Fe/H] plane. 

Our simulated Milky Way accretes a sim$\_$Sequoia analogue at a similar time as the sim$\_$Enceladus, with the two satellites accreted from the same dominant filamentary structure but from opposite sides. This is an interesting result as it shows that the progenitor galaxies of GE and Sequoia are separate entities that seem to come from different origins in this simulation, which makes contrast with other studies that suggest that Sequoia was part of GE \citep[e.g.][]{koppelman2020, naidu2021, amarante2022}. 

\begin{figure}
\begin{tabular}{c}
{\includegraphics[width=3.2in, trim={0.5cm 1.48cm 0.5cm 0.5cm}, clip]{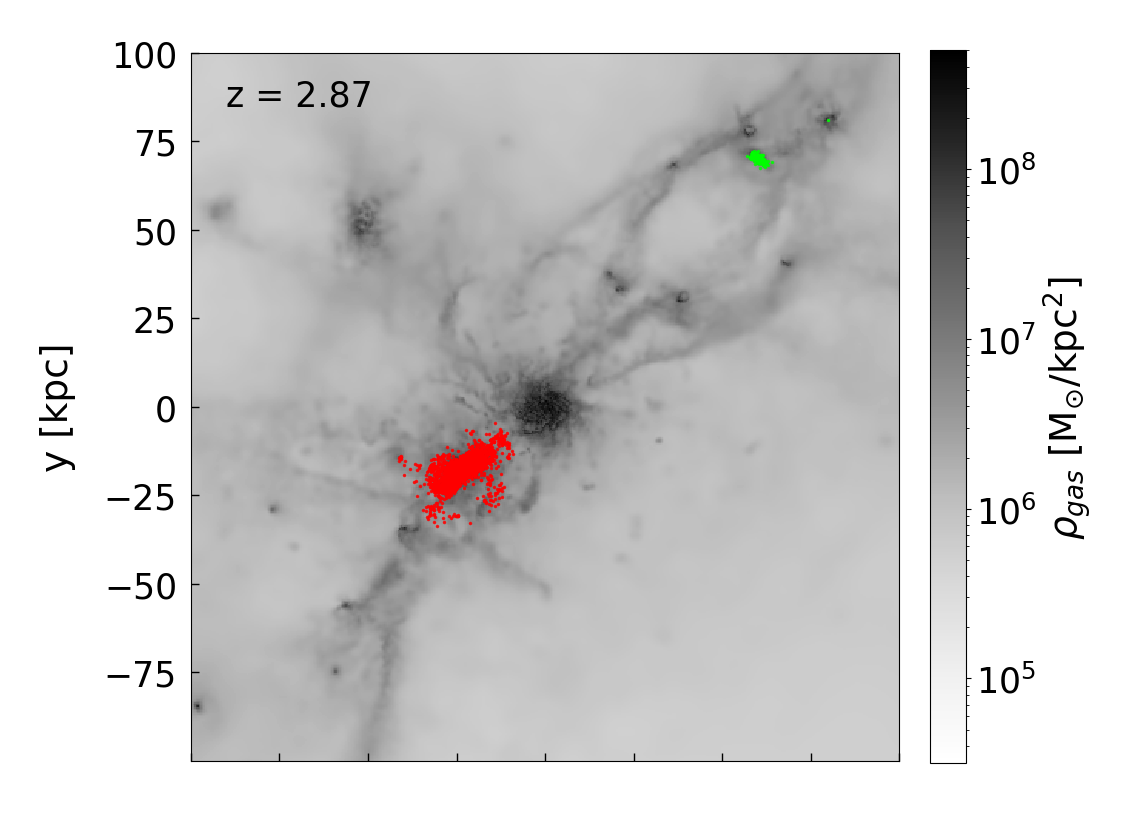}}\\
{\includegraphics[width=3.2in, trim={0.5cm 0.5cm 0.5cm 1.05cm}, clip]{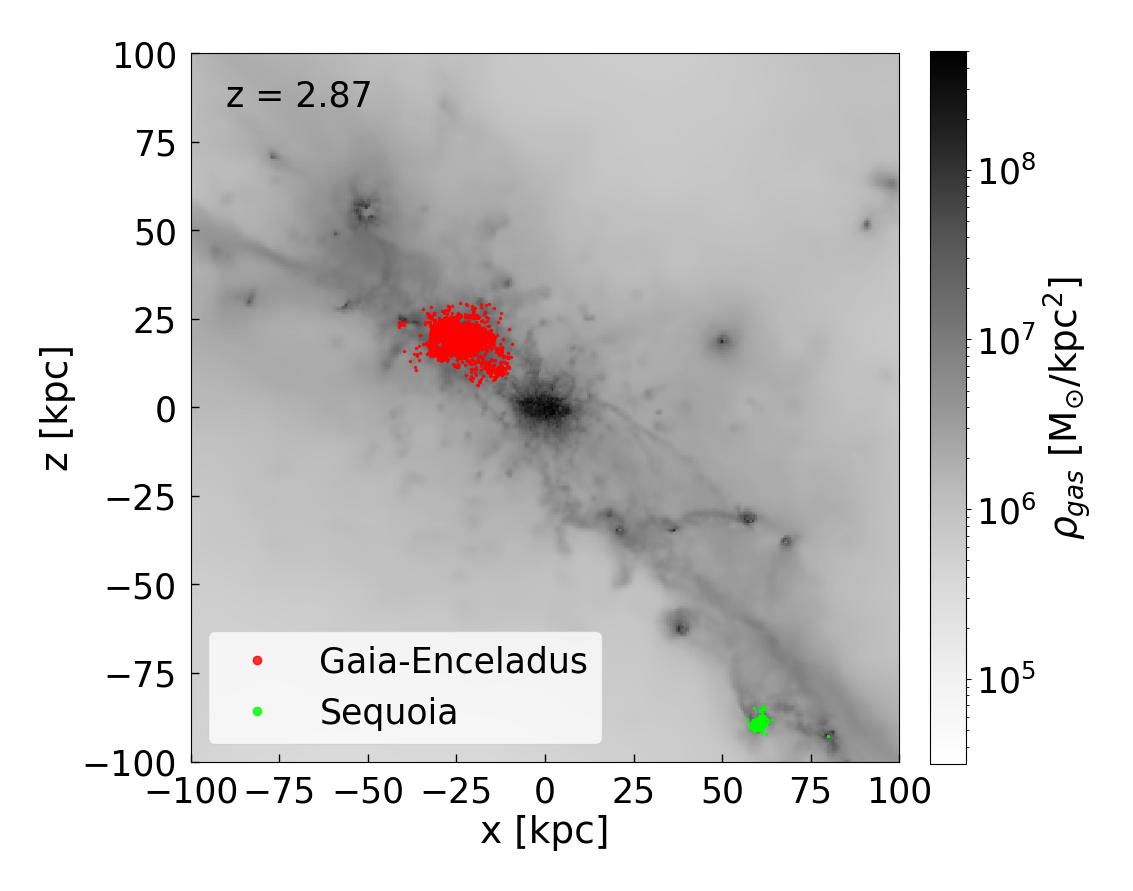}} 
\end{tabular}
\caption{The position of sim$\_$Enceladus (red) and sim$\_$Sequoia (green) stars at the time prior to the accretion of sim$\_$Enceladus. The background is the gas density, and shows that the two satellites are both accreted from the most prominent filament, but from opposite directions. The central galaxy is aligned face on (top panel) and edge on (bottom panel).}
\label{fig:filaments} 
\end{figure}

Additionally, in a recent work presented by \cite{an_beers2022}, a nearly in-plane collision between GE and the main galaxy is proposed to explain the stellar distribution of this merger within the Milky Way. In our simulation, however, sim$\_$Enceladus made its infall from above the galactic plane (as we can see on the bottom panel of Fig.~\ref{fig:filaments}), and yet, this analogue has proven to be well-matched to GE properties. What is perhaps the most interesting result in our study is the very close resemblance of these two simulated accreted satellites  to the observed ones in the \textit{j$_z$}-Energy plane (Fig.~\ref{fig:IoM}), including the significant retrograde motion of sim$\_$Sequoia.  Importantly, this series of accretion events occur within a fully CDM cosmological setting in which we simultaneously form a central galaxy with thin and thick discs which also have been shown to closely match the observed Milky Way \citep{brook2020chemical}. 

In addition, regions in the \textit{j$_z$}-Energy plane defined by different groups of stars are shown to be consistent with respect to their merging time, i.e. there is a clear relation between the time of accretion and the final average energy of the stars. This relation has already been discussed in previous studies \citep[e.g.][]{amorisco2017, pfeffer2020, horta2023}.

In chemical abundance space,  sim$\_$Sequoia matches very well the observed Sequoia \citep{matsuno2022}, with the downturn in  [$\alpha$/Fe]-[Fe/H] at a lower value of [Fe/H] than sim$\_$Enceladus, reflecting its lower mass. 

In our simulation, sim$\_$Sequoia is accreted slightly after sim$\_$Enceladus, although there is little difference in the average age of the stars of the two accreted satellites. We are not claiming that this reflects the real order of accretion of these satellites, but we do want to emphasise that the average ages of their stars depend on the details of their star formation history prior to their accretion, as well as on any ongoing star formation that may occur during accretion. Hence, phenomena such as the evolution of the gas density of mergers due to tidal stripping and ram pressure before and during the accretion, or the growth of the main galaxy with time, may play an important role in this process, and may depend on orbital parameters as well as mass. 

\section{Conclusions} \label{sec:concl}
We have presented the merger history of a fully cosmological Milky Way analogue simulation that has analogues of three key early merger events in the observed Milky Way, namely Kraken, Gaia-Enceladus and Sequoia. The main results of this paper are summarised as follows: 

\begin{itemize}
    \item The three mergers characterised in this work are found to strongly match their observed properties, including their distributions in kinematic phase space and in IoM space. It is also observed a relation in which, early accreted stars tend to show lower energies and small rotation, whilst later structures have high energies and more prograde or retrograde motion.
    \item We find very good agreement with the observed properties of the Sequoia merger, including its characteristic retrograde rotation in IoM space and its chemical abundances, being clearly separate from those of Gaia-Enceladus, which has higher values of [Fe/H] and [O/Fe] in average. This is a result that supports the high-precision abundances observed in \cite{matsuno2022}.
    \item The results for our Kraken-like merger have close similarity with those from \cite{orkney2022}, in terms of being chemically indistinguishable from the main galaxy at the time of infall of this structure. Therefore, it overlaps populations of old in-situ stars. This merger shows large dispersion in all three velocity components and resides in the innermost part of the galactic halo, being strongly bound to the system as well. 
    \item In this simulation, sim$\_$Enceladus and sim$\_$Sequoia have similar average ages: the latter is slightly younger because it is accreted later than the former. These analogues infall along the same dominant filament but from opposite directions. This provides a scenario where the two galaxies are distinct and did not infall together as part of a group.
\end{itemize}

We emphasise again that, presumably due to the similarities in merger history, this simulation has been shown to match a large range of chemo-dynamic properties of the real  Milky Way including the chemo-dynamical properties of the thick and thin discs and stellar halo, as well as reproducing the remnants of a Gaia-Enceladus analogue. Simulated galaxies of similar mass using the same model but with different merger histories do not match nearly as well the properties of the Milky Way. Further, simulations using the same physical model has been shown to match a large range of scaling relations.  The success of this model therefore provides support that, with the combined efforts of observations and simulations,  we are arriving at a detailed understanding of the evolutionary history of our Galaxy within a cold dark matter cosmological framework.

\section*{Acknowledgements}
CB is supported by the Spanish Ministry of Science and Innovation
(MICIU/FEDER) through research grant PID2021-122603NB-C22. RG acknowledges financial support from the Spanish Ministry of Science and Innovation (MICINN) through the Spanish State Research Agency, under the Severo Ochoa Program 2020-2023 (CEX2019-000920-S), and support from an STFC Ernest Rutherford Fellowship (ST/W003643/1). DK acknowledges the support of UK Research and Innovation (EP/X031756/1) and the STFC Grant (ST/W001136/1).

\textit{Software:} \textsc{pynbody} \cite[]{pynbody}, \textsc{numpy} \cite[]{numpy}, \textsc{matplotlib} \cite[]{matplotlib}, \textsc{seaborn} \cite[]{seaborn}.

\section*{Data Availability}
The data underlying this article is available upon request, \href{chbrook@ull.edu.es}{chbrook@ull.edu.es}.

\vspace{.0cm}\bibliographystyle{mn2e}
\bibliography{archive}


\label{lastpage}

\end{document}